\documentclass[aps,prd,floatfix,superscriptaddress,amsmath,amssymb,twocolumn,nofootinbib]{revtex4-1}

\usepackage[colorlinks=true,citecolor=blue,linkcolor=blue,breaklinks=true]{hyperref}
\usepackage{amsmath}
\usepackage[T1]{fontenc}
\usepackage{amsfonts}
\usepackage{amssymb}
\usepackage{bbold}
\usepackage{epsfig}
\usepackage{graphicx}
\usepackage{bm}
\usepackage{array}
\usepackage{xspace}
\usepackage{listings}
\usepackage{color}
\usepackage{float}
\usepackage{multirow}
\usepackage[normalem]{ulem}

\usepackage{tikz,xcolor,hyperref}

\definecolor{lime}{HTML}{A6CE39}
\DeclareRobustCommand{\orcidicon}{\hspace{-1mm}
	\begin{tikzpicture}
	\draw[lime, fill=lime] (0,0) 
	circle [radius=0.16] 
	node[white] {{\fontfamily{qag}\selectfont \tiny \,ID}};
	\draw[white, fill=white] (-0.0525,0.095) 
	circle [radius=0.007];
	\end{tikzpicture}
	\hspace{-3mm}
}

\foreach \x in {A, ..., Z}{\expandafter\xdef\csname orcid\x\endcsname{\noexpand\href{https://orcid.org/\csname orcidauthor\x\endcsname}
			{\noexpand\orcidicon}}
}

\begin{document}

\title{\textit{R}-process $\beta$-decay neutrino flux from binary neutron star mergers and collapsars}

\author{Yu An\orcidA{}}
\email{anyu2017gfo@gmail.com}
 \affiliation{Key Laboratory of Dark Matter and Space Astronomy, Purple Mountain Observatory, Chinese Academy of Sciences, Nanjing 210023}
 \affiliation{School of Astronomy and Space Science, University of Science and Technology of China, Hefei 230026}
 
\author{Meng-Ru Wu\orcidB{}}%
\email{mwu@gate.sinica.edu.tw}
\affiliation{Institute of Physics, Academia Sinica, Taipei 11529}
\affiliation{Institute of Astronomy and Astrophysics, Academia Sinica, Taipei 10617}
\affiliation{Physics Division, National Center for Theoretical Sciences, Taipei 10617}

\author{Gang Guo\orcidC{}}
\affiliation{School of Mathematics and Physics, China University of Geosciences, Wuhan 430074}

\author{Yue-Lin Sming Tsai\orcidD{}}
\affiliation{Key Laboratory of Dark Matter and Space Astronomy, Purple Mountain Observatory, Chinese Academy of Sciences, Nanjing 210023}
 \affiliation{School of Astronomy and Space Science, University of Science and Technology of China, Hefei 230026}

\author{Shih-Jie Huang\orcidE{}}
\affiliation{Department of Physics, National Taiwan University, Taipei 10617}

\author{Yi-Zhong Fan\orcidF{}}
\affiliation{Key Laboratory of Dark Matter and Space Astronomy, Purple Mountain Observatory, Chinese Academy of Sciences, Nanjing 210023}
 \affiliation{School of Astronomy and Space Science, University of Science and Technology of China, Hefei 230026}

\date{\today}

\begin{abstract}
This study investigates the antineutrinos production by $\beta$-decay of $r$-process nuclei in two astrophysical sites that are capable of producing gamma-ray bursts (GRBs): binary neutron star mergers (BNSMs) and collapsars, which are promising sites for heavy element nucleosynthesis. 
We employ a simplified method to compute the $\beta$-decay $\bar\nu_e$ energy spectrum and consider a number of different representative thermodynamic trajectories for $r$-process simulations, each with four sets of $Y_e$ distribution. 
The time evolution of the $\bar\nu_e$ spectrum is derived for both the dynamical ejecta and the disk wind for BNSMs and collapsar outflow, based on approximated mass outflow rates.  
Our results show that the $\bar\nu_e$ has an average energy of 
approximately 3 to 9~MeV, with a high energy tail of up to 20 MeV.  
The $\bar\nu_e$ flux evolution is primarily determined by the outflow duration, and can thus remain large for $\mathcal{O}(10)$~s and $\mathcal{O}(100)$~s for BNSMs and collapsars, respectively. 
For a single merger or collapsar at 40~Mpc, the $\bar\nu_e$ flux is $\mathcal{O}(10-100)$~cm$^{-2}$~s$^{-1}$, indicating a possible detection horizon up to $0.1-1$~Mpc for Hyper-Kamiokande. 
We also estimate their contributions to the diffuse $\bar\nu_e$ background, 
and find that both sources should only contribute subdominantly to the diffuse background when compared to that expected from core-collapse supernovae.
\end{abstract}

\maketitle

\section{Introduction}
\label{sec:intro}
Half of the elements heavier than iron are thought to be synthesized via the rapid neutron capture process ($r$-process)~\cite{Cowan:2019pkx,Arcones:2022jer}. 
During the $r$-process, seed nuclei are driven close to the neutron drip line by successive neutron captures that operate much faster than the $\beta$ decays.    
The typically required astrophysical conditions for $r$-process—high neutron number density and short dynamical expansion timescales—may be satisfied in some specific  explosions of core-collapse supernovae (CCSNe)~\cite{Qian:1996xt,Winteler+12,Mosta:2017geb,Siegel:2018zxq,Fischer:2020xjl} and binary neutron star mergers (BNSMs)~\cite{1999ApJ...525L.121F, Goriely:2011vg, Fernandez:2013tya, Just:2014fka}. 
While the detection of optical/infrared transient AT2017gfo, the associated kilonova of GW170817, provided the first piece of convincing evidence of $r$-process in BNSMs~\cite{LIGOScientific:2017vwq, LIGOScientific:2017ync, Kasen:2017sxr, Shibata:2017xdx, Metzger:2019zeh, Watson:2019xjv}~ (see also \cite{Tanvir:2013pia,Yang:2015pha,Jin:2016pnm} for earlier kilonova signals in support of this scenario), it remains unclear whether rare types of supernova are needed as additional $r$-process sources. 
Among those, collapsars, which originate from rapidly rotating massive stars that undergo gravitational collapse to form black holes surrounded by massive accretion disks, are plausible progenitors of certain Type Ic supernovae associated with long gamma-ray bursts (LGRBs)~\cite{Woosley:1993wj, MacFadyen:1998vz, Woosley:2006fn, Kumar:2014upa}. 
Ref.~\cite{Siegel:2018zxq} found that collapsar disks may have conditions similar to the BNSM remnant disks and can potentially be another important or even dominant $r$-process site. 
However, this possibility remains highly debated from both the theoretical~\cite{Miller:2019mfl, Zenati:2020ncf,Fujibayashi:2020jfr,Just:2022fbf,Fujibayashi:2022xsm} and observational~\cite{Macias:2019oxw, Bartos:2019twj, Brauer:2020hty, Lee:2022ijg,Anand:2023ujd} points of view.

In addition to the electromagnetic and gravitational wave radiation, neutrinos are produced copiously in BNSMs and collapsars.  
Neutrino emissions from these sources include the low energy ones with $E_\nu\sim\mathcal{O}(10)$~MeV emitted by thermal processes from the central remnants~\cite{Kyutoku:2017wnb,Cusinato:2021zin,Wei:2019hpd} as well as potentially the high energy (HE) ones with $E_\nu\sim\mathcal{O}(1-10^3)$~TeV~\cite{Meszaros:2001ms, Razzaque:2004yv, Murase:2013ffa, Senno:2015tsn, Tamborra:2015fzv, Guo:2022zyl, Fang:2017tla, Decoene:2019eux, Kimura:2019ipr, Gottlieb:2021pzr}. 
The direct detection of these neutrinos from live events is very challenging due to their weakly interacting nature and searches have thus far returned null results~\cite{ANTARES:2017bia, Super-Kamiokande:2018dbf, Baikal-GVD:2018cya, ANTARES:2018bmu, IceCube:2020xks, IceCube:2022rlk, IceCube:2023atb}.
However, a future detection of them may provide important clues to conditions and properties deep inside the sources. 
Moreover, both the thermal and the HE neutrinos may contribute significantly as diffuse sources relevant to the search for diffuse neutrinos from astrophysical sources.

As the $\beta$ decays constantly happen during and after the $r$-process, the electron antineutrinos ($\bar\nu_e$) are naturally produced in any $r$-process sites. 
The amount of produced $\bar\nu_e$ from a single $r$-process site can be estimated as
$N_{\bar\nu_e}\simeq \Delta Y_e M_{\rm ej} / m_u$, where $M_{\rm ej}$ is the total ejecta mass, $m_u$ is the atomic mass unit, and $\Delta Y_e\equiv Y_e^f-Y_e^i$ is the difference between the final and initial electron number fraction per baryon $Y_e\equiv N_e/N_b$ with $N_{e(b)}$ 
denoting total number of electrons (baryons) in the ejecta.
For $\Delta Y_e\simeq 0.2$ and $M_{\rm ej}\simeq 1 M_\odot$, we have $N_{\bar\nu_e}\simeq 2.5\times 10^{56}$, which can be comparable to the amount of $\bar\nu_e$ emitted by a CCSN. 
These $\beta$-decay $\bar\nu_e$ are expected to have an average energy of $\langle E_{\bar\nu_e}\rangle\sim 5$~MeV, which is related to the average mass difference between of $\beta$-decay parent and daughter nuclei during the $r$-process~\cite{Guo:2022zyl,Chen:2023mn}. 
They are generally being emitted at radii $\gtrsim \mathcal{O}(10^3)$~km from the expanding ejecta and do not affect the dynamics and composition of the system. 
However, their annihilation with the HE neutrinos, if produced in for instance mildly magnetized jets of collapsars, can potentially leave important imprints on the energy spectrum and flavor ratio of HE neutrinos, which are viable candidates of the diffuse flux being observed by the IceCube~\cite{Guo:2022zyl}.

Reference~\cite{Chen:2023mn} recently estimated the time evolution of average energy and luminosity of $r$-process $\beta$-decay $\bar\nu_e$ emitted from materials ejected promptly (e.g., the dynamical ejecta) in a BNSM by defining a radiation efficiency that accounts for the average energy carried by $\bar\nu_e$ from $\beta$ decays. 
It was found that the luminosity evolution of $\bar\nu_e$ follows that of radioactive heating rate, which takes roughly a constant value for $\mathcal{O}(1)$~s and later decreases following $\propto t^{-1.3}$ after the $r$-process ends~\cite{Metzger:2010sy, Korobkin:2012uy}. 
However, outflows from the accretion disks that can last for $\mathcal{O}(1-10)$~s, are expected to dominate that of dynamical ejecta for BNSMs, and can even last longer for $\mathcal{O}(10-100)$~s in the case of collapsars. 
Consequently, one expects a very different temporal evolution of the corresponding $\bar\nu_e$ flux emitted from that derived in Ref.~\cite{Chen:2023mn}. 

In this work, we examine in greater detail the $\bar\nu_e$ emission from the BNSM dynamical ejecta as well as the disk winds from both BNSMs and collapsars for individual events, and their contributions to the diffuse $\bar\nu_e$ flux. 
We first use a simplified framework to estimate the $\bar\nu_e$ energy spectrum from each $\beta$-decay nucleus in Sec.~\ref{sec:beta_decay_sim}.
We follow the abundance evolution of $r$-process nuclei in materials with parameterized expansion history and $Y_e$ distributions that take different average $Y_e$ values, using an $r$-process nuclear reaction network detailed in Sec.~\ref{sec:r_process}.  
In Sec.~\ref{sec:neutrino_flux}, we compute the emitted $\bar\nu_e$ energy spectrum as well as the evolution of their flux and average energy for different outflow components of BNSMs and collapsars.  
We then estimate their resulting contributions to the diffuse neutrino background and compare that with the diffuse SN neutrino background in Sec.~\ref{sec:diffuse_neutrino} and summarize our findings in Sec.~\ref{sec:conclusion}.

\section{$\beta$-decay neutrino calculation}
\label{sec:beta_decay_sim}
For a nucleus $^A_Z X$ with mass number $A$ and atomic number $Z$, its ground-state to ground-state $\beta$ decay proceeds as 
\begin{align}
    ^A_Z X \rightarrow & ^A_{Z+1} Y + e^- + \bar{\nu}_e, 
\end{align}
to produce a daughter nucleus $^A_{Z+1}Y$, an electron $e^-$, and an antineutrino $\bar\nu_e$.
For an allowed transition, the differential decay rate that produces an $e^-$ with kinetic energy $T_e$  
can be expressed as~\cite{Fermi:1934hr, Schopper:1969jkp}
\begin{equation}
    \begin{aligned}
    \frac{d\lambda}{dT_e} 
        =&\frac{G_F^2 |M^0_{fi}|^2}{2\pi^3 }F(Z_{d}, T_{e})C(T_e)(Q-T_{e})^2 \\
    &\sqrt{T_{e}^2 + 2T_{e} m_{e}}(m_{e}+T_{e}),
    \label{eq:allowed_beta}
\end{aligned}
\end{equation}
where $\lambda$ is the decay rate, $G_F$ is the Fermi constant, $M^{0}_{fi}$ is the transition matrix element from the initial state to the final state with the superscript $0$ denoting the allowed transition, $F(Z_d, T_e)$ and $C(T_e)$ are Fermi function and shape factor, $Z_d=Z+1$, and $Q\equiv m(^A_Z X)-m(^A_{Z+1} Y)-m_e$. 
The Fermi function $F(Z_d, T_e)$ is a correction factor for the Coulomb interaction~\cite{Fermi:1934hr, 1985JPhG...11..359V}, and could be written as 
\begin{equation}
\begin{split}
    F(Z_d,T_e)=
    &\frac{2(1+S)}{\Gamma(1+2S)^2}(2\sqrt{T_{e}^2 + 2T_{e} m_e}\rho_d)^{2S-2}e^{\pi\eta}\\
    &|\Gamma(S+i\eta)|^2,
    \label{fermi_function}
\end{split}
\end{equation}
where $S = \sqrt {1-\alpha^2 Z_d^2}$, $\alpha$ is the fine-structure constant, $\Gamma$ denotes the Gamma function, $\rho_d \simeq R_d$ is the radius of daughter nucleus, which could be estimated as $R_d = 1.3A^{1/3}$~fm, $\eta=Z_d e^2/\hbar v$ for $\beta^{-}$ decay, and $v$ is the electron velocity.  
The shape factor $C(T_e)$ describes the angular momentum taken by the electron-neutrino pair, and it simply equals to 1 for allowed transitions. 

Although Eq.~\eqref{eq:allowed_beta} is only valid for allowed transitions between ground states, we utilize it and make several simplifications to approximately calculate the $\beta$-decay energy spectra of $e^-$ and $\bar\nu_e$ for all nuclei involved in $r$-process. 
We assume that for each nucleus, the aggregated $e^-$ spectrum can be approximated by the same shape as given in Eq.~\eqref{eq:allowed_beta}. 
To roughly take into account the energy loss due to the deexcitation $\gamma$ for branches where the daughter nucleus are in excited states, we define an effectively reduced $Q$-value by $\tilde Q=P\cdot Q$ and take $P=0.9$. 
With these assumptions, the normalized $\bar\nu_e$ spectrum from a nucleus $^A_Z X$ can be written as 
\begin{equation}
 \begin{split}
     f_{(Z,A)}(E_{\bar\nu_e})=
     &
     K
     F(Z_d, E_{\bar{\nu}_e})E_{\bar{\nu}_e}^2(\tilde{Q} - E_{\bar{\nu}_e}+m_e)\\
     &\times \sqrt{(\tilde{Q} - E_{\bar{\nu}_e})^2 + 2(\tilde{Q} - E_{\bar{\nu}_e}) m_{e}},
\label{eq:fnuebar}
\end{split}
\end{equation}
where $K$ is a normalization constant such that $\int_0^{\tilde Q}dE_{\bar\nu_e}f_{(Z,A)}(E_{\bar\nu_e})=1$. 
Note that in writing down Eq.~\eqref{eq:fnuebar}, we first replace $Q$ by $\tilde Q$ in Eq.~\eqref{eq:allowed_beta} and then use $E_{\bar\nu_e}=\tilde Q-T_e$. 
The $Q$ values are computed based on experimentally known nuclear masses from \cite{Wang:2021xhn} and the theoretically predicted masses from the finite-range droplet model (FRDM)~\cite{Moller:1993ed} for the rest. 
We compare in Appendix~\ref{appendix:beta_decay_spectra} the shape of the corresponding $e^-$ spectra using different values of $P$ for several isotopes whose $\beta$-decay $e^-$ spectra were measured experimentally.

\begin{table*}[t]
    \centering
    \setlength{\tabcolsep}{0.27cm}
    \begin{tabular}{c | c c | c c}
        \hline
        \hline
        \multirow{2}{*}{Model} & \multicolumn{2}{c|}{Trajectory parameters} & {M$_{\rm ej}$ [M$_{\odot}$]} & {Mass outflow rate}\\
        \cline{2-3}
        & $s [k_B/$baryon$]$ & $\tau [$ms$]$ \\ 
        \hline
        BNSM-dyn. A & 1  & 0.2  & 0.01 & prompt \\
        BNSM-dyn. B & 20 & 2 & 0.01 & prompt  \\
        BNSM-disk   & 15 & 20 & 0.05 & secular; see Eq.~\eqref{eq:outflow_rate} \\
        collapsar   & 15 & 20 & 0.50 & secular; see Eq.~\eqref{eq:outflow_rate}\\
        \hline
        \hline
    \end{tabular}
    \caption{
    The adopted parameters for different models considered in this work. 
    The expansion history of the density and temperature of the ejecta are modeled analytically with two parameters, the entropy $s$ and the expansion timescale $\tau$~\cite{Lippuner:2015gwa}. 
    $M_{\rm ej}$ is the assumed total ejecta mass for each model.
    The BNSM dynamical ejecta (BNSM-dyn. A and BNSM-dyn. B) are assumed to be promptly ejected while the BNSM-disk and collapsar mass outflow rates are assumed to follow Eq.~\eqref{eq:outflow_rate}. 
    For all cases, we adopt four different sets of $Y_e$ distributions with central values $Y_{e,c}=0.15$, 0.25, 0.35, and 0.45 (see text for detail). 
    }
    \label{tab:model}
\end{table*}

With Eq.~\eqref{eq:fnuebar}, we can then easily compute the total $\bar\nu_e$ emissivity (defined as number of $\bar\nu_e$, $N_{\bar\nu_e}$, emitted per unit mass $M$ per unit time per unit energy) at any location inside the ejecta by summing over all $\beta$-decay nuclei 
\begin{equation}
    \frac{dN_{\bar{\nu}_e}}{dM dE_{\bar{\nu}_e} dt}=\frac{1}{m_u}
    \sum_{(Z,A)} f_{(Z,A)}(E_{\bar\nu_e}) \lambda_{(Z,A)} Y({Z,A}), 
    \label{eq:nu_emiss}
\end{equation}
where $\lambda_{(Z,A)}$ is the net $\beta$-decay rate of the parent nucleus and 
$Y({Z,A})=n_{({Z,A})}/(\rho/m_u)$ denotes the abundance of a nuclear species $^A_Z X$ with $n_{({Z,A})}$ the corresponding number density and $\rho$ the total mass density of a parcel inside the ejecta. 
For $\lambda_{(Z,A)}$, we take experimentally measured values from Ref.~\cite{Audi:2017asy} when available and adopt theoretical values from~\cite{PhysRevC.67.055802} otherwise. 
The abundance $Y({Z,A})$ will be computed with the $r$-process nuclear reaction network described in the next section.  

\section{Expansion model and $r$-process}
\label{sec:r_process}
\begin{figure*}[ht]
    \centering
    \includegraphics[width=\textwidth]{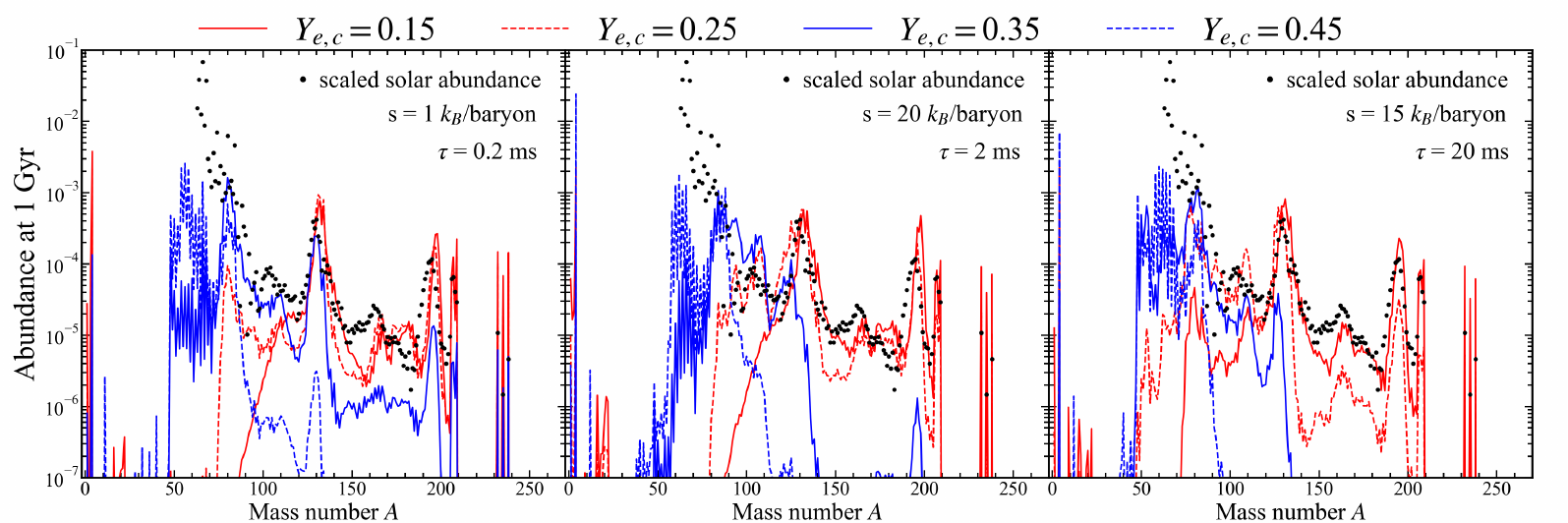}
    \caption{
    Nuclear abundance, $Y(A)$ as a function of mass number $A$ at 1~Gyr for four different sets of $Y_e$ distributions with central value $Y_{e,c}=0.15$, $0.25$, $0.35$, and $0.45$ described in the main text. 
    The left, middle, and right panels show results obtained using trajectories with expansion dynamical timescale $\tau=0.2$~ms, $\tau=2$~ms and $\tau=20$~ms, entropy $s=1$~$k_B$, $s=20$~$k_B$ and $s=15$~$k_B$ per nucleon, which are named as Case A, B and C in the text, respectively. 
    Also shown are the rescaled solar $r$-process abundances (black dots).  
    }
    \label{fig:r-process}
\end{figure*}
We use an established 
nuclear reaction network code used in~\cite{deJesusMendoza-Temis:2014owk,Wu:2016pnw,Wu:2018mvg} to compute the $r$-process nucleosynthesis.
The network consists of over 7000 isotopes from neutron up to $^{110}_{313}$Ds and includes all relevant nuclear reactions including the neutron capture $(n,\gamma)$ and its inverse photodissociation $(\gamma, n)$ rates, the charged particle reactions, $\beta$-processes ($\beta^{-}$ decays, $\beta^{+}$ decays, electron captures), $\alpha$ decays, and fissions (neutron-induced, spontaneous, $\beta$-delayed).
Details regarding the nuclear physics inputs to the network can be found in Ref.~\cite{deJesusMendoza-Temis:2014owk}.

The network calculation also requires the expansion history of ejecta as inputs. 
Instead of taking trajectories directly from hydrodynamical simulations, we adopt a parametrized model proposed in Ref.~\cite{Lippuner:2015gwa}, which characterizes the density evolution of the ejecta as 
\begin{align}
\rho(t) = \left\{\begin{array}{ll}
\rho_0 e^{-t/\tau} & \text{if $t \leq 3\tau$}, \\
\rho_0 \left(\dfrac{3\tau}{et}\right)^3 & \text{if $t \geq 3\tau$},
\end{array}\right. 
\label{eq:rho_evol}
\end{align}
where $\tau$ is the expansion timescale and $\rho_0$ is the initial mass density at time $t=0$. 
We assume three representative cases: Case A with $\tau=0.2$~ms and initial entropy $s=1$~$k_B$/nucleon, where $k_B$ is the Boltzmann constant, to represent the fastest-expanding and low-entropy ejecta, Case B with $\tau=2$~ms and initial entropy $s=20$~$k_B$/nucleon for the slower-expanding and higher-entropy ejecta, and Case C with $\tau=20$~ms and initial entropy $s=15$~$k_B$/nucleon for the slowest expanding and intermediate-entropy ejecta. 
The choice of these values are guided by the simulation data used in Refs.~\cite{deJesusMendoza-Temis:2014owk,Wu:2016pnw,George:2020veu} and will be associated with different ejecta components in the next section.
For a given initial temperature $T(t=0)$ and the entropy, the value of $\rho_0$ can be computed through the equation of state of \cite{Timmes:2000eos}. 
The subsequent temperature evolution of the ejecta $T(t)$ is 
assumed to follow the relation that $T^3/\rho$ is a constant\footnote{We note that this assumption implies that we have assumed the entropy of the ejecta remain approximately as a constant. Including the feedback of nuclear energy release on the entropy evolution may affect the quantitative evolution of nucleosynthesis, especially for initially low-entropy case ($s\lesssim 1$)~\cite{deJesusMendoza-Temis:2014owk}. 
However, we expect that the qualitative features related to $\bar\nu_e$ emission, especially for the higher entropy cases, will not be affected significantly.}. 

The initial neutron richness condition, which can be related to the initial value of $Y_e$, is the most pivotal parameter that determines the $r$-process outcome. 
Since recent hydrodynamical simulations of BNSMs and collapsars generally predict a broad distribution of $Y_e$ ranging in between $\sim 0.01$ to $0.5$, we take the same approach as in Ref.~\cite{Wu:2018mvg} by assuming a normalized distribution of $Y_e$ in ejecta, characterized by a Gaussian function with a central value $Y_{e,c}$ and a half width $\Delta Y_e$. 
For each expansion model, we take 50 initial $Y_e$ values ranging from 0.01 to 0.5 (with an interval of 0.01) to perform the $r$-process calculations.
For each calculation, we evolve the nuclear abundance evolution $Y_{(Z,A)}(t)$ starting from a temperature $T_0=10$~GK where the initial $Y_{(Z,A)}(t=0)$ are determined by the condition of nuclear statistical equilibrium (NSE) until $t=10^{18}$~s. 
For any quantities $\mathcal{Q}(t)$ related to a specific ejecta model, we then sum over the corresponding normalized $Y_e$ distribution $G(Y_{e,i};Y_{e,c},\Delta Y_e)$ as
\begin{equation}
    \mathcal{Q}(t) = \sum_{i=1}^{50} \mathcal{Q}(t, Y_{e,i}) G(Y_{e,i};Y_{e,c}, \Delta Y_e), 
    \label{eq:normalization}
\end{equation}
where $\sum_{i=1}^{50} G(Y_{e,i};Y_{e,c},\Delta Y_e) =1$.
We have taken four different values of $Y_{e,c}=0.15, 0.25, 0.35, 0.45$ with a fixed $\Delta Y_e=0.04$ for Case A, B and C.

Figure~\ref{fig:r-process} shows the resulting abundance distribution as a function of the mass number, $Y(A)=\sum_Z Y(Z,A)$, at the time of 1~Gyr for all cases described above. 
Clearly, those with lower $Y_{e,c}$ values produce distributions toward heavier nuclei as they have larger initial neutron richness for $r$-process to proceed further. 
With $Y_{e,c}=0.15$, both the second ($A\sim 130)$ and third ($A\sim 195$) peaks as well as the actinides, are produced and qualitatively agree with the scaled solar $r$-process abundance distribution taken from~\cite{2008ARA&A..46..241S}. 
However, the first peak nuclei ($A\sim 80)$ are largely underproduced.   
For $Y_{e,c}=0.25$, all three peaks are produced in Case C, while Case A and B do not produce much of the first-peak nuclei. 
For $Y_{e,c}=0.35$, Case A is still capable of producing a robust second peak, which is barely produced in Case B and C.
As for $Y_{e,c}=0.45$, all three cases present a strong iron peak~($A\sim 56)$ and a first peak.
In general, the low-entropy and fast-expanding Case A gives rise to more enhanced production of heavier nuclei compared to Case B and C with higher entropy and slower expansion. 
The major reason is that the trajectories in Case A generally have higher average mass numbers in seed nuclei than Case B and C before the onset of the $r$-process, due to their lower entropy. 
This, in turn, leads to a higher average mass number after the $r$-process ends in Case A, and correspondingly the formation of heavier nuclei. 

\section{Neutrino flux from $r$-process sources}
\label{sec:neutrino_flux}

In this section, we combine the results derived in Secs.~\ref{sec:beta_decay_sim} and \ref{sec:r_process} with different mass outflow models that represent different ejecta components from BNSM and collapsar events, to derive the expected $\bar\nu_e$ energy spectra and fluxes from them.

\begin{figure*}[t]
    \centering
    \includegraphics[width=\textwidth]{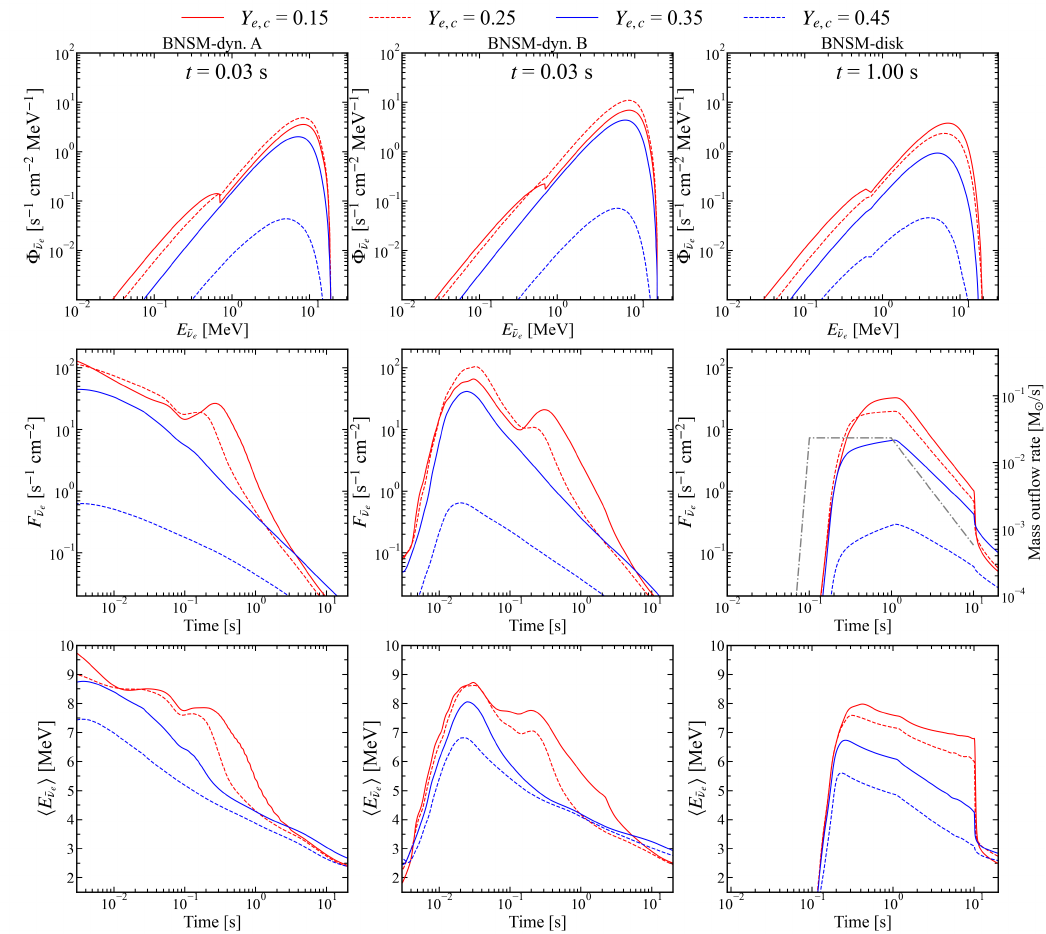}
    \caption{
    The $r$-process $\bar\nu_e$ energy spectrum $\Phi_{\bar\nu_e}$ (top row), the time evolution of the $\bar\nu_e$ flux, $F_{\bar\nu_e}$ (middle row), and the time evolution of the $\bar\nu_e$ average energy, $\langle E_{\bar\nu_e} \rangle$ (bottom row), from the model of BNSM-dyn.~A (left column), BNSM-dyn.~B (middle column), and BNSM-disk (right column), respectively. 
    The $\bar\nu_e$ energy spectra shown in the top panels are taken at 0.03~s, 0.03~s, and 1.0~s, approximately at when the corresponding fluxes are near their peak values. 
    Different lines in each panel represent cases with different $Y_{e,c}$ values of 0.15, 0.25, 0.35, and 0.45, whose $r$-process abundances are shown in Fig.~\ref{fig:r-process}. 
    In the right middle panel, the BNSM-disk wind mass outflow rate is shown by the gray dashed-dotted line. 
}
    
    \label{fig:neutrino_bns}
\end{figure*}

\subsection{Binary neutron star merger}
Recent simulations of BNSMs suggest two main types of ejecta: the prompt ejecta that are dynamically ejected within timescales of $\sim\mathcal{O}(10)$~ms and the long-term outflow driven from the merger remnant consisting of a central BH or a hypermassive neutron star (HMNS) with an accretion disk, which can persist for $\sim\mathcal{O}(1-10)$~s~\cite{Shibata:2019wef}. 
The dynamical ejecta consist of materials that are tidally disrupted as well as those squeezed out from the contact interface when the two neutron stars collide. 
The ejecta mass of either component, $M_{\rm dyn}$, may range between $10^{-4}$~$M_\odot$ to $10^{-2}$~$M_\odot$, depending on the binary mass ratio as well as the nuclear EoS. 
Because the expansion history of different dynamical ejecta components can be different, we examine the $\bar\nu_e$ emission from them with expansion parameters given by Case A and B separately as model BNSM-dyn. A and BNSM-dyn. B, respectively (see Table~\ref{tab:model}). 

For the dynamical ejecta, since their mass ejection timescale $\sim\mathcal{O}(10)$~ms is typically much shorter than the $r$-process time scale $\sim\mathcal{O}(1)$~s, we neglect the time dependence of mass ejection and assume that all ejecta are ejected promptly at the same time for simplicity. 
Under this approximation, the $r$-process $\bar\nu_e$ spectrum flux (per unit energy per unit time per unit area) at a distance $D$ far away from the source can be easily computed as 
\begin{equation}
    \Phi_{\bar\nu_e,\rm dyn}(E_{\bar{\nu}_e},t) = \frac{M_{\rm dyn}}{4 \pi D^2} \frac{dN_{\bar{\nu}_e}}{dM dE_{\bar{\nu}_e} dt},
    \label{eq:flux_dyn}
\end{equation}
where the expression of $dN_{\bar{\nu}_e}/(dM dE_{\bar{\nu}_e} dt)$ was given in Eq.~\eqref{eq:nu_emiss}. 
Using the above equation, we also compute the total $\bar\nu_e$ number flux $F_{\bar\nu_e,\rm dyn} \equiv \int dE_{\bar\nu_e} \Phi_{\bar\nu_e,\rm dyn}(E_{\bar\nu_e},t)$ as well as their average energy $\langle E_{\bar\nu_e} \rangle_{\rm dyn}\equiv \left ( \int dE_{\bar\nu_e} E_{\bar\nu_e} \Phi_{\bar\nu_e,\rm dyn} \right) / \left( \int dE_{\bar\nu_e} \Phi_{\bar\nu_e,\rm dyn} \right ) $. 

The left and middle panels in Fig.~\ref{fig:neutrino_bns} show $\Phi_{\bar\nu_e,\rm dyn}(E_{\bar{\nu}_e},t_p)$ as functions of $E_{\bar\nu_e}$ at a specific time $t_p$, 
as well as $F_{\bar\nu_e,\rm dyn}(t)$ and $\langle E_{\bar\nu_e} \rangle_{\rm dyn} (t)$ as functions of time, for model BNSM-dyn.~A and model BNSM-dyn.~B, respectively, for all four neutron-rich conditions of different $Y_{e,c}$. 
We take $M_{\rm dyn}=10^{-2}$~$M_\odot$ and $D=40$~Mpc, and choose $t_p=0.03$~s, 
which roughly corresponds to the peak time in $F_{\bar\nu_e,\rm dyn}(t)$ for model BNSM-dyn.~B. 
Comparing both models, the resulting $\bar\nu_e$ spectra take similar shapes for cases with the same $Y_{e,c}$ at $t_p$. 
Both spectra peak at $E_{\bar\nu_e}\simeq 4-10$~MeV with higher energy tails extending up to $\sim 20$~MeV.
On the lower energy end, the spectra for cases with $Y_{e,c}=0.15$ show a subleading component below $\sim 1$~MeV, which is due to the contribution from the decay of free neutrons that are still abundantly present during the $r$-process.  
On the other hand, the evolution of the total number flux and the average energy behave very differently at earlier time in both models for
$t\lesssim t_p$.  
model BNSM-dyn.~A has $F_{\bar\nu_e,\rm dyn}$ and $\langle E_{\bar\nu_e} \rangle_{\rm dyn}$ that decrease slowly in the beginning while for Model
BNSM-dyn.~B, $F_{\bar\nu_e,\rm dyn}$ and $\langle E_{\bar\nu_e} \rangle_{\rm dyn}$ first increase until $\sim t_p$ and then decrease over time.
The major difference at the early time is that for BNSM-dyn.~A with low entropy, the seed nuclei are present initially at $T=10$~GK at very neutron-rich side. 
This permits fast enough $\beta$ decays at the beginning of the simulation and thus generates large $\bar\nu_e$ flux at early times. 
However, for BNSM-dyn.~B, the formation of seed nuclei only takes place at lower temperature
due to the higher entropy, which occurs at a later time $\gtrsim 0.005$~s. 
Moreover, they are with lower mass numbers, less neutron-rich, and thus lower $\beta$-decay rates. 
Thus, the $\bar\nu_e$ flux is initially low and only increases when $r$-process proceeds and moves nuclei to heavier ones that are more neutron rich. 
We also note that for the two lower $Y_e$ cases ($Y_{e,c}=0.15$ and $0.25$), there is a bump around $t\sim 0.3$~s for both Model A and B, which is related to the nucleosynthesis flow reaching the region above the third peak. 

These figures also show that for cases with $Y_{e,c}=0.15, 0.25$, and $0.35$, the peak values of $F_{\bar\nu_e,\rm dyn}$ 
only differ by $\sim$ a factor of $2-3$, reaching up to $\sim 30-100$~s$^{-1}$~cm$^{-2}$ for the chosen fiducial source distance $D=40$~Mpc away from Earth, which can become comparable to or larger than the $\bar\nu_e$ flux from the irreducible background of diffuse supernova neutrino background (DSNB) (see Sec.~\ref{sec:diffuse_neutrino}).
For $Y_{e,c}=0.45$, since there are little $r$-process, the $\bar\nu_e$ flux is also smaller than the other cases by more than one order of magnitude. 
For the average energy, $3$~MeV$\lesssim \langle E_{\bar\nu_e} \rangle_{\rm dyn}\lesssim 9$~MeV during the relevant duration, which is broadly consistent with values estimated in Refs.~\cite{Guo:2022zyl,Chen:2023mn}.

Besides the dynamical ejecta, subsequent mass outflows can be launched from the post-merger remnant disk on a time scale of $\mathcal{O}(1-10)$~s.  
Numerical simulations show that up to $\sim 20\%-50\%$ of the initial remnant disk mass, up to $0.1$~$M_\odot$, can be ejected mainly by viscous heating, after the thermal neutrinos emission from the disk becomes inefficient~\cite{Fujibayashi:2022ftg,Fahlman:2022jkh}~(see Appendix~\ref{appendix:thermal_neutrino} for the estimation of thermal $\bar\nu_e$ emission from the post-merger remnant). 
The $Y_e$ distribution of the disk outflow can be greatly influenced by the lifetime of the HMNS~\cite{Lippuner:2017bfm, Fujibayashi:2022ftg, Just:2023wtj}. 
For cases with the prompt formation of BHs, the disk outflows may have average $Y_e$ values around $0.2-0.25$, while neutrino irradiation from longer-lived HMNSs over $\mathcal{O}(10-100)$~ms can raise the average outflow $Y_e$ to values above 0.3.  
Since the major part of the disk outflows tends to have slower expansion velocity and higher entropy, we only examine the expansion parameter set case C for the disk outflows, named Model BNSM-disk (see Table~\ref{tab:model}). 

Unlike the dynamical ejecta, the disk winds are keep being launched for a time duration longer than the $r$-process time scale. 
Thus, a mass outflow rate as a function of time $\dot M_{\rm disk}(t)$ is needed to compute the associated $r$-process $\bar\nu_e$ emission. 
We parameterize $\dot M_{\rm disk}(t)$ based on recent MHD remnant disk simulations in Ref.~\cite{Fahlman:2022jkh}.  
The outflow rate $\dot M_{\rm disk}(t)$ generally has a steeply rising phase,  followed by a nearly constant plateau, before it declines with a power-law of $\propto t^{-1.6}$. 
Specifically, we take 
\begin{align}
\dot M_{\rm disk}(t) = \left\{\begin{array}{lll}
A_0 t^{\alpha} & \text{$t_0 < t < t_1$}, \\
C_0 & \text{$t_1 \leq t \leq t_2$}, \\
B_0 t^{\beta} & \text{$t_2 < t < t_3$},
\end{array}\right. 
\label{eq:outflow_rate}
\end{align}
with $\alpha = 15$, $t_0 = 10$~ms after the merger to approximate the rising phase until $t_1 = 10^{-1}$~s. 
From $t_1$ to $t_2 = 1$~s, $\dot M_{\rm disk}$ takes a constant value of $C_0$, after which it decays with $\beta = -1.6$ until $t_3=10$~s. 
The values of $A_0$, $C_0$ and $B_0$ are determined by requiring the continuity of $\dot M_{\rm disk}(t)$ at $t_1$ and $t_2$ as well as having $\int_{t_0}^{t_3} dt \dot M_{\rm disk}(t) = M_{\rm disk}=5\times 10^{-2}$~$M_\odot$, which is the assumed total disk outflow mass. 

Summing over the contribution of different mass shells ejected at different times,  the resulting disk $r$-process $\bar\nu_e$ flux spectrum (at $D$ far away from the source) can be computed by 
\begin{equation}
    \Phi_{\bar\nu_e,\rm disk}(E_{\bar{\nu}_e},t) = 
    \frac{1}{4\pi D^2}
    \int_{t_0}^{t_3} dt' 
    \left [ {\dot M}_{\rm disk}(t')\Theta(t-t') \frac{dN_{\bar{\nu}_e}(t-t')}{dM dE_{\bar{\nu}_e} dt} \right ] ,
    \label{eq:flux_disk}
\end{equation}
where $\Theta(t-t')$ is the Heaviside function. 

\begin{figure*}[ht]
    \centering
    \includegraphics[width=\textwidth]{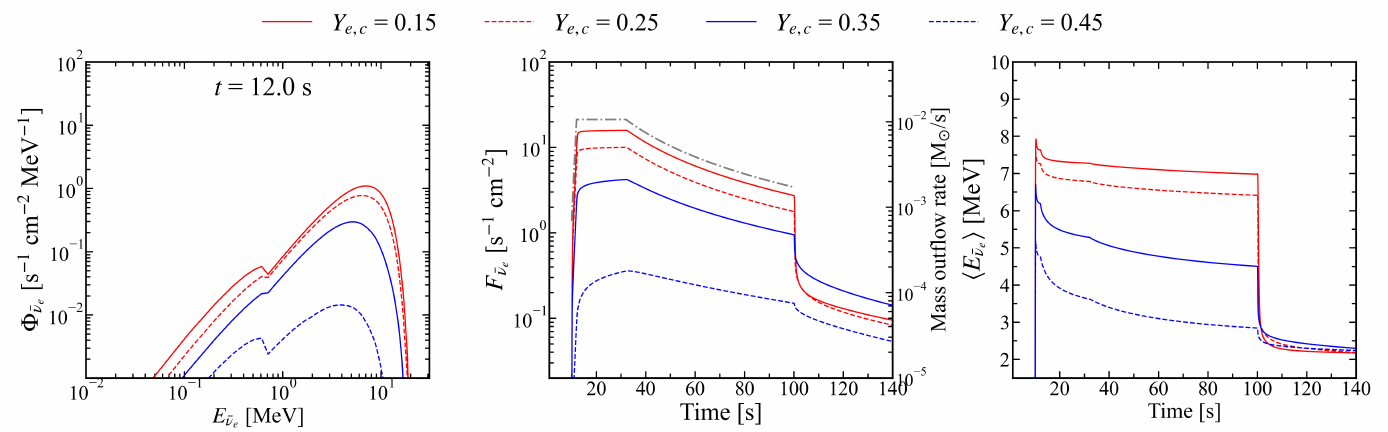}
    \caption{
    The $r$-process $\bar\nu_e$ energy spectrum $\Phi_{\bar\nu_e}$ (left panel), the time evolution of the $\bar\nu_e$ flux, $F_{\bar\nu_e}$ (middle panel), and the time evolution of the $\bar\nu_e$ average energy, $\langle E_{\bar\nu_e} \rangle$ (right panel) from the collapsar outflow model. 
    The $\bar\nu_e$ energy spectrum is taken at 12.0~s.
    In the middle panel, the mass outflow rate is shown by the gray dashed-dotted line. 
    }
    \label{fig:flux_collapsar}
\end{figure*}

We show $\Phi_{\bar\nu_e,\rm disk}(E_{\bar\nu_e},t=1~{\rm s})$  as well as the corresponding $F_{\bar\nu_e,\rm disk}(t)$ and $\langle E_{\bar\nu_e} \rangle_{\rm disk}(t)$ (similarly defined as before) in the right column of Fig.~\ref{fig:neutrino_bns} for Model BNSM-disk with different $Y_{e,c}$. 
The shape of the $\bar\nu_e$ energy spectrum at the time when $F_{\bar\nu_e,\rm disk}$ peaks for all cases are similar to those derived for dynamical ejecta. 
However, both $F_{\bar\nu_e,\rm disk}$ and $\langle E_{\bar\nu_e} \rangle_{\rm disk}$ evolve completely different from the dynamical ejecta cases where prompt mass ejection was assumed.  
Clearly, the shape of $F_{\bar\nu_e,\rm disk}$ roughly follows that of ${\dot M}_{\rm disk}$, with a rapid increase in the beginning, a nearly plateau phase for $\sim 1$~s, and then followed by the decline phase, due to the continuous contribution of $\bar\nu_e$ emission from the secular outflow.
For $Y_{e,c}=0.15, 0.25$ and $0.35$, their plateau values of $F_{\bar\nu_e,\rm disk}$ can reach $\sim 5-40$~s$^{-1}$~cm$^{-2}$ for $D=40$~Mpc, only slightly lower than the peak values of the dynamical ejecta. 
For the case with the highest $Y_{e,c}=0.45$, the flux is once again lower than other cases by more than one order of magnitude. 
As for the average $\bar\nu_e$ energy, it can remain relatively stable 
with $\langle E_{\bar\nu_e}\rangle_{\rm disk} \sim 5-9$~MeV for most of the time when $t\lesssim t_3$ when matter outflow is nonzero.
This is because the $r$-process in outflows launched later keeps producing a substantial amount of $\bar\nu_e$ with higher energy.
Only when $\dot M_{\rm disk}$ stops, $\langle E_{\bar\nu_e}\rangle_{\rm disk}$ drops to much smaller values $\lesssim 3$~MeV\footnote{Note that the very steep drop in $F_{\bar\nu_e,\rm disk}$ and $\langle E_{\bar\nu_e}\rangle_{\rm disk}$ are related to the sudden termination of $\dot M_{\rm disk}$. In reality, the transition will be slightly more smooth.
}. 
\subsection{Collapsar} 

Reference~\cite{Siegel:2018zxq} recently proposed that outflows from the accretion disk formed in collapsars can possibly have neutron-rich conditions similar to that obtained in BNSMs during phases when the mass accretion rate is larger than $\sim 10^{-3}$~$M_\odot$~s$^{-1}$, and may lead to a massive amount of outflows of $\lesssim 1$~$M_\odot$ that are enriched by $r$-process. 
Follow-up theoretical works~\cite{Miller:2019mfl, Fujibayashi:2020jfr, Zenati:2020ncf} have not yet found similar conditions as given in Ref.~\cite{Siegel:2018zxq}. 
Early attempts to search for $r$-process evidence based on analyzing light curves associated with Type Ic SNe~\cite{Anand:2023ujd} or consideration based on $r$-process abundances in metal-poor stars~\cite{Macias:2019oxw, Bartos:2019twj, Brauer:2020hty} also suggested that collapars may not be the dominating $r$-process sources. 
However, large uncertainties remain in both theory and observation and it is yet too early to exclude collapsars as possible $r$-process sites. 

Given these large uncertainties in $r$-process conditions, we again take model collapsar with four different $Y_{e,c}$ values to investigate the $r$-process $\bar\nu_e$ production in collapsar outflows.  
Given that the outflows are driven from the disk wind, but at a much longer timescale~\cite{Just:2022fbf}, we take the same shape of $\dot M_{\rm disk}$ as in Eq.~\eqref{eq:outflow_rate}, but with $t_0 = 10$~s, $t_1= 12$~s, $t_2= 32$~s and $t_3=100$~s. 
The coefficients $A_0$, $B_0$, and $C_0$ are determined by having a total outflow mass $M_{\rm col}=0.5$~$M_\odot$.
The resulting $\bar\nu_e$ flux spectrum can also be similarly computed using Eq.~\eqref{eq:flux_disk}. 

Fig.~\ref{fig:flux_collapsar} shows the resulting $\bar\nu_e$ energy spectrum at $t=12$~s as well as the evolution of their total number flux and the average energy, again, for a source at $D=40$~Mpc. 
The obtained results are generally similar to those from the BNSM disk winds—with similar peak flux values that sustain for a much longer time duration due to the involved longer outflow timescale.

\subsection{Detection prospects}
Based on the $\bar\nu_e$ fluxes computed for individual BNSM or collapsar events, we estimate the prospect of detection as follows. 
Taking the inverse beta decay (IBD) as the detection channel for $\bar\nu_e$, the event rate in a water Cherenkov detector containing $N_p$ target protons can be roughly estimated as\footnote{Here we have ignored the flavor conversions of neutrinos, which can suppress the $\bar\nu_e$ flux at Earth by an $\mathcal{O}(1)$ factor. The precise value of the suppression factor depends on the yet-unknown neutrino mass ordering as well as the detailed density profile of the ejecta.}
\begin{equation}
    \mathcal{R}(t) \approx N_p \bar\sigma_{\rm IBD} F(t),
\end{equation}
where $\bar\sigma_{\rm IBD}\approx 9.5\times 10^{-42}$~cm$^2\times[\langle E_{\bar\nu_e}\rangle / (10~{\rm MeV})]^2$. 
Taking a typical peak flux $F(t)\simeq 30$~cm$^{-2}$~s$^{-1}\times [(40~{\rm Mpc})/D]^2$ (see Figs.~\ref{fig:neutrino_bns} and \ref{fig:flux_collapsar}), $N_p\sim 1.3\times 10^{34}\times [M_T/(200~{\rm kton})]$, one obtains
\begin{equation}
    \mathcal{R} \approx 0.6~{\rm s}^{-1}
    \times [(100~{\rm kpc})/D]^2 \times [M_T/(200~{\rm kton})].
\end{equation}
For the BNSM disk wind and collapsar outflow whose $r$-process $\bar\nu_e$ emission duration are $\mathcal{O}(10)$~s and $\mathcal{O}(100)$~s, one may expect to detect $\gtrsim\mathcal{O}(1)$ events for $D\lesssim \mathcal{O}(300)$~kpc and $D\lesssim\mathcal{O}(1)$~Mpc, respectively, for individual events assuming a detector size close to that of Hyper-Kamiokande~\cite{Ruggeri:2023btm}. 
For dynamical ejecta, it would require a Galactic event with $D\lesssim\mathcal{O}(10)$~kpc to have $\gtrsim\mathcal{O}(1)$ IBD events from the $r$-process $\bar\nu_e$, consistent with what estimated in Ref.~\cite{Chen:2023mn}. 

Note that the numbers reported above only serve as a qualitative measure. 
A complete analysis to quantify the detection of $r$-process neutrinos will require taking into account 
several factors, including e.g., the neutrino oscillations, the detector response, and various background contributions from reactor antineutrinos, atmospheric antineutrinos, neutral-current quasielasitic scattering, and DSNB~\cite{Kyutoku:2017wnb}. 
Moreover, thermal neutrinos that are coproduced from the same source (BNSM remnant or collapsar disk) will also need to be taken into account. 
For thermal neutrinos, we expect that they can overshadow most of the $r$-process $\bar\nu_e$ flux from the BNSM dynamical ejecta due to their expected larger fluxes at times $\lesssim \mathcal{O}(100)$~ms~\cite{Cusinato:2021zin}. 
However, except for the situation where a long-lived  $\gtrsim \mathcal{O}(1)$~s central remnant is obtained in a BNSM, the emissions of thermal neutrinos from the BH--disk in BNSM or in collapsar disk should become subdominant by the time when the $r$-process $\bar\nu_e$ are produced, since the disk outflows are only launched when neutrino cooling of the disk becomes insignificant, as illustrated in Fig.~\ref{fig:thermal_nue_flux} in Appendix~\ref{appendix:thermal_neutrino}.
Although the probability of having a BNSM or a collapsar within the horizon of a Hyper-Kamiokande-like detector is small, if such a rare event occurs, the (non-)detection of the $r$-process neutrinos from the disk outflow will provide direct diagnostics to $r$-process nucleosynthesis therein.

\section{Diffuse neutrino background} \label{sec:diffuse_neutrino}
In addition to the possible direct detection of $\bar\nu_e$ from nearby individual $r$-process events, their accumulated flux from \emph{all past sources} manifests as a diffuse $r$-process neutrino background (D$r$NB) and can also be evaluated in the same way as how the DSNB are typically computed~\cite{Ando:2004hc,Beacom:2010kk, Suliga:2022ica}. 
Below, we use results derived in the previous section to compute the potential flux of D$r$NB from BNSMs and collapsars, compare them to the DSNB flux 
and diffuse thermal $\bar\nu_e$ flux from BNSMs. 

We write the present (redshift $z=0$) D$r$NB flux $\Phi_{\nu,{\rm D}r{\rm NB}}(E_{\bar\nu_e})$ by summing over the contribution from all past $r$-process $\bar\nu_e$ emitting sources up to $z_{\rm max}$ as

\begin{equation}
    \Phi_{\nu,{\rm D}r{\rm NB}}(E_{\bar\nu_e}) = 
    c \int_0^{z_{\rm max}} dz 
    (1 + z)R(z)\frac{dt}{dz} \frac{dN_{\bar\nu_e}(E^{\prime}_{\bar\nu_e})}{dE^{\prime}_{\bar\nu_e}},
    \label{eq:diffuse_flux}
\end{equation}
where $E^{\prime}_{\bar\nu_e}$ = (1 + $z$)$E_{\bar\nu_e}$ is the $\bar\nu_e$ energy at redshift $z$ that will be locally observed as energy $E_{\nu}$,  $R(z)$ is the occurring rate of the $\bar\nu_e$ sources (BNSMs or collapsars) per comoving volume at $z$, and  $dN_{\bar\nu_e}$/$dE_{\bar\nu}$ is the emitted $\bar\nu_e$ number spectrum for a single event. 
In Eq.~\eqref{eq:diffuse_flux}, the factor $dt/dz$ from the standard $\Lambda$CDM cosmological model is given as 
\begin{equation}
    \frac{dt}{dz} = -\frac{1}{H_0(1 + z)\sqrt{\Omega_m (1 + z)^3 + \Omega_{\Lambda}}},
    \label{eq:t_redshift}
\end{equation}
where we have taken $\Omega_m$ = 0.3, $\Omega_{\Lambda}$ = 0.7 and $H_0$ = 70 km s$^{-1}$ Mpc$^{-1}$. 

\begin{figure}[tbp]\centering
    \includegraphics[width=1.0\columnwidth]{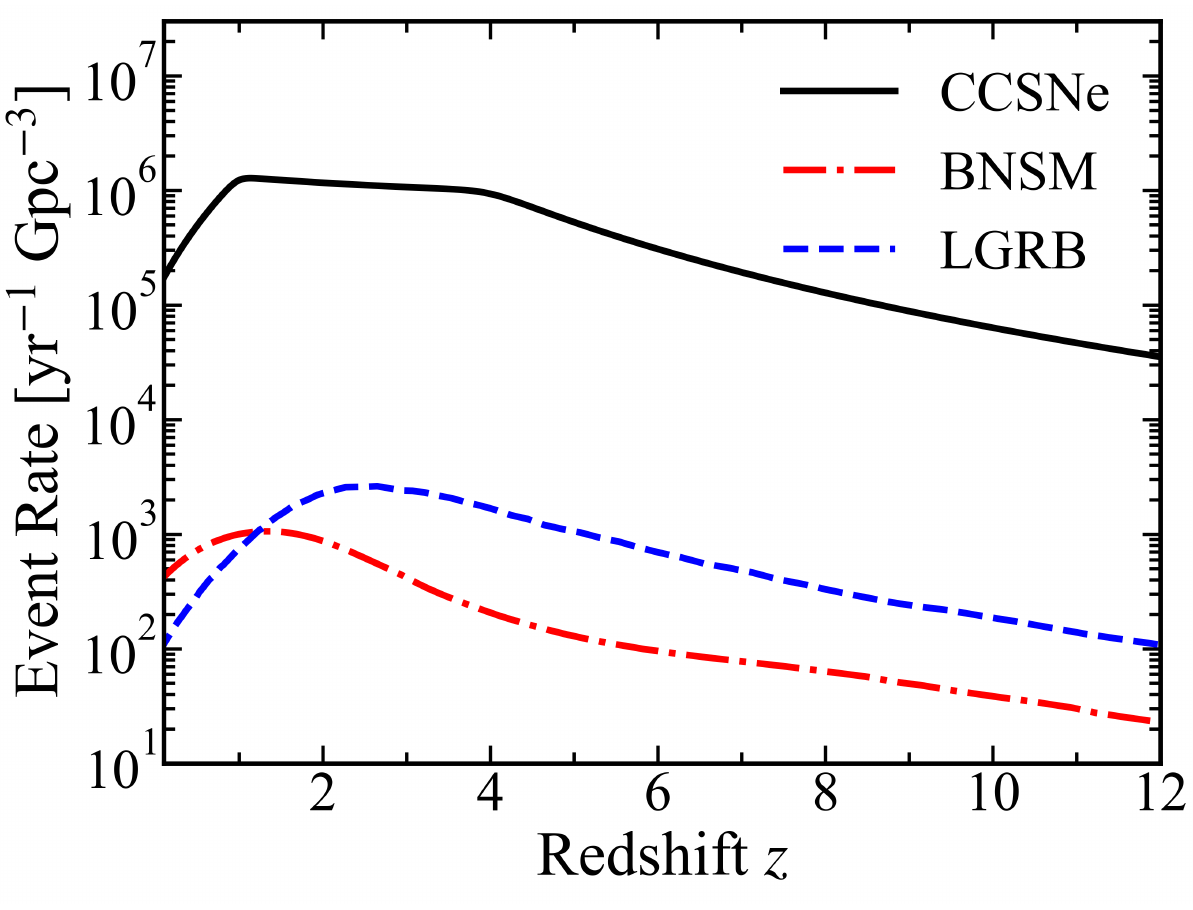}
    \caption{
    The adopted event rate per comoving volume $R(z)$ as a function of redshift $z$ for BNSM (the CC15$\alpha$5 model in \cite{Santoliquido:2020axb}, red dash-dotted line), CCSNe~\cite{Horiuchi:2008jz} (black solid line), and collapsars (blue shaded area). 
    For the collapsar rate, we take the LGRB rate~\cite{Ghirlanda:2022edk} (blue dash line) as the lower limit.
        } 
    \label{fig:event_rate}
\end{figure}

For the redshift-dependent event rate $R(z)$ of BNSMs, we adopt the result obtained by Ref.~\cite{Santoliquido:2020axb}, which performed studies of the cosmological merger rate density for compact mergers based on a population synthesis model that takes into account the impact of various physical parameters for the common envelope (CE), natal kicks, CCSNe, initial mass function (IMF), metallicity, star formation rate (SFR), and mass transfer efficiency, on different binary compact star merger scenarios. 
As a representative case, we use the derived BNSM 
rate from their model CC15$\alpha$5 (see the red solid curve in Figure 12 of Ref.~\cite{Santoliquido:2020axb})\footnote{This paper found that models using a CE parameter $\alpha_{\rm CE} \geq 2$ (high CE ejection efficiency) and a low natal kick are required to satisfy the local rate determined by the LIGO-Virgo collaboration.
}.
For collapsars, besides as sources of LGRBs whose rates can be inferred, they may also be responsible for the low-luminosity GRBs or even choked GRBs, whose rates are highly uncertain and can be higher than that of LGRBs. 
However, it is unlikely that all of them (on average) eject $\sim 1$~$M_\odot$ of
$r$-process materials.
This is because simple estimates suggested that on average $\lesssim 0.3$~$M_\odot$ of $r$-process materials are ejected assuming that LGRBs are the main source for Milky Way's $r$-process inventory~\cite{Siegel:2018zxq}.
Thus, we take the cosmic LGRB rate from Ref.~\cite{Ghirlanda:2022edk} as the collapsar rate capable of contributing to the D$r$NB.
We plot all these rates along with the SN rates from Ref.~\cite{Horiuchi:2008jz} assuming the Salpeter IMF (see Appendix~\ref{appendix:neutrino_spectra} for details).   
The adopted BNSM rate peaks at $z\simeq 1.2$ and is generally three orders of magnitude smaller than that of SNe. 
For the collapsar rate, it peaks at a higher redshift around $z\simeq 3$, and can be potentially as large as one-tenth of the SN rate. 

\begin{figure}[t]
    \centering
    \includegraphics[width=1.0\columnwidth]{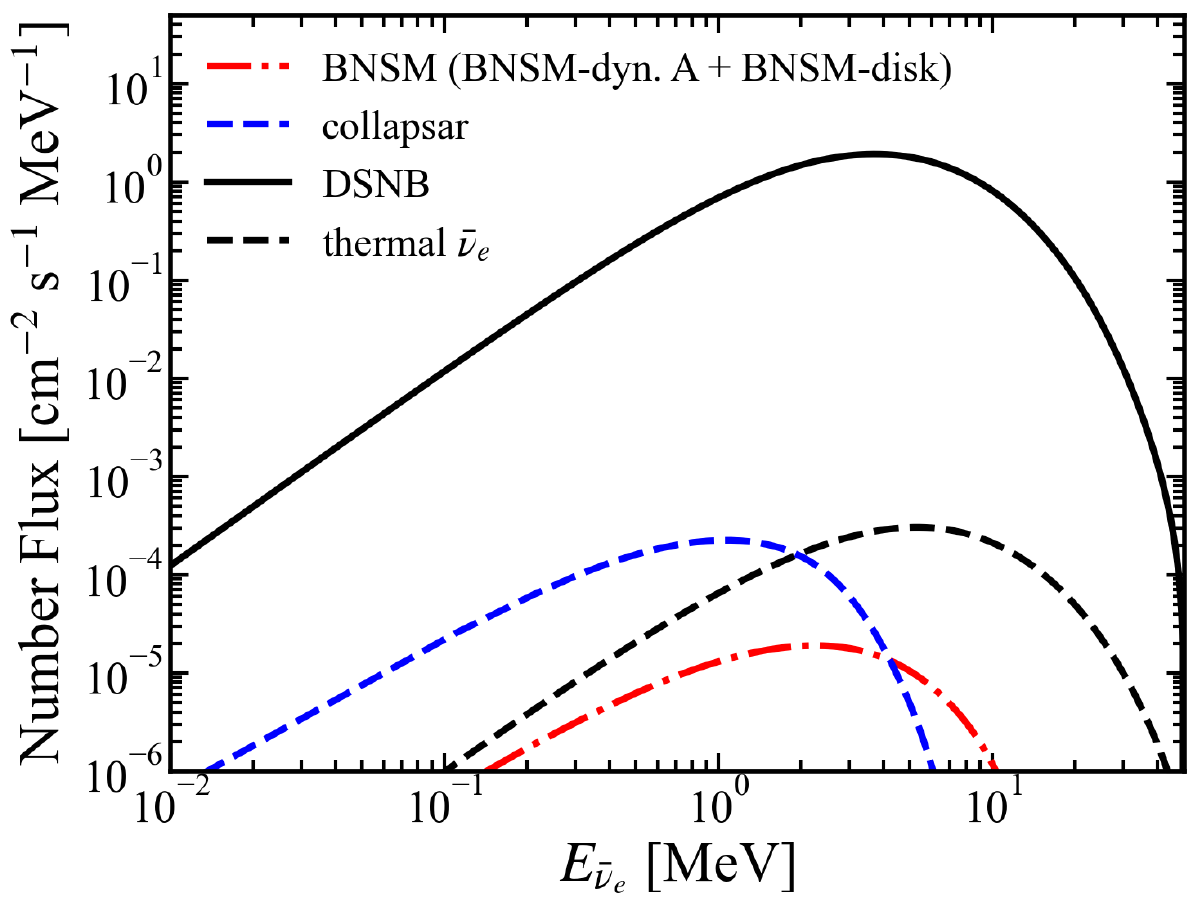}
    \caption{Diffuse $r$-process $\beta^-$-decay neutrino flux from BNSM (red dash-dotted line) and collapsars (blue shaded area). 
    Also shown are the the DSNB flux and diffuse thermal BNSM $\bar\nu_e$ flux (black solid line and black dash line) for comparison.
    The range of the collapsar flux is associated with the assumed uncertain event rate shown in Fig.~\ref{fig:event_rate}. 
    For the BNSM flux, it includes both the contribution from the dynamical ejecta (model BNSM-dyn.~A)
    and the disk wind (model BNSM-disk).
    We use the $Y_e$ distribution with central value of $Y_{e,c}=0.25$ for both the BNSM and collapsar cases. 
    }
    \label{fig:diffuse_flux}
\end{figure}

Given the above $R(z)$, we then sum both the $\bar\nu_e$ emitted from the 
model BNSM-dyn.~A and the model BNSM-disk with $Y_{e,c}=0.25$ as the source spectrum $dN_{\bar\nu_e}(E_{\bar\nu_e})/dE_{\bar\nu_e}$ for all BNSMs\footnote{Here we integrate the previously computed $dN_{\bar\nu_e}/( dE_{\bar\nu_e} dt)$ over a time period of 50~s and 140~s, for BNSM and collapsar outflows, respectively. We checked that the chosen durations are long enough to contain most of $\bar\nu_e$ emitted in both cases.}. 
For collapsars, we take the collapsar disk outflow result discussed earlier also with $Y_{e,c}=0.25$.
Fig.~\ref{fig:diffuse_flux} shows the derived D$r$NB from both BNSMs and collapsars.
For comparison, we also show the DSNB flux computed by following Ref.~\cite{Horiuchi:2008jz} with an assumed $\bar\nu_e$ source spectrum characterized by an effective temperature parameter  $T_{\nu_e}=5$~MeV~\cite{Beacom:2010kk}, 
as well as the diffuse thermal $\bar\nu_e$ flux from BNSMs computed in Appendix~\ref{appendix:thermal_neutrino}.
Clearly, the D$r$NB flux from BNSMs is roughly 4 to 5 orders of magnitude smaller than that of DSNB, which is due to both the lower source rate and the smaller amount of emitted $\bar\nu_e$ per event. 
Comparing to the diffuse thermal neutrinos from the same source, the D$r$NB flux is also roughy one order of magnitude smaller.
For the D$r$NB from collapsars, their flux may be as large as that of BNSM thermal neutrinos, but are still 2 to 3 orders of magnitude smaller than the DSNB within the considered energy range\footnote{Note that here we do not consider the thermal component from collapsars for the diffuse flux. However, it is very likely that this component also dominates the $r$-process one from collapsars, but still smaller than the DSNB; see e.g., the C0(GRB1) curve in Fig.~5 of Ref.~\cite{Schilbach:2018bsg}.}.
Thus, we conclude that the detection of D$r$NB is unlikely and the presence of them will hardly affect the upcoming detection of DSNB.

\section{Summary and discussions}

We have estimated the $\bar\nu_e$ energy spectrum originated from the decay of unstable neutron-rich nuclei produced during the $r$-process nucleosynthesis by utilizing a simplified treatment of $\beta$ decays.  
Taking three parametrized expansion histories of ejecta, each of which representing different components of the BNSM ejecta or collapsar outflows, we have also computed the expected $r$-process $\bar\nu_e$ flux and average energy from BNSMs and collapsars by following the detailed evolution of nuclear abundances with a nuclear reaction network for different assumed $Y_e$ distributions.  
Our results show that the main feature of the $\bar\nu_e$ energy spectrum during the $r$-process are qualitatively similar for all different scenarios, characterized by an averaged energy of $\sim 5-9$~MeV with an extended tail up to $\sim 20$~MeV.
The time evolution of the $\bar\nu_e$ flux for the BNSM dynamical ejecta peaks at $\lesssim 1$~s followed by a power-law decay profile. 
However, for the BNSM disk wind and the collapsar outflows whose ejection duration are longer than the $r$-process timescale, we found that the $\bar\nu_e$ flux evolution closely follows the mass outflow rate of the ejecta. 
There, the flux can remain large for $\mathcal{O}(1-10)$~s and $\mathcal{O}(10-100)$~s for the BNSM disk wind and collapsar outflow, respectively. 
For all scenarios, we found that the value of the $\bar\nu_e$ flux peak are within a factor of a few so long as a non-negligible fraction of second $r$-process peak nuclei is produced, corresponding to our choice of averaged $Y_e$ values below 0.35.
However, for higher averaged $Y_e$ value at e.g., 0.45, the $\bar\nu_e$ flux are suppressed substantially. 

Assuming that the BNSM dynamical ejecta, BNSM disk wind, and collapsar outflows have masses of $0.01$~$M_\odot$, $0.05$~$M_\odot$, and $0.5$~$M_\odot$, we estimated that to detect the $r$-process $\bar\nu_e$ from BNSM disk wind and collapsars, it would likely require a BNSM or a collapsar to occur at a distance of $\lesssim\mathcal{O}(300)$~kpc or $\lesssim\mathcal{O}(1)$~Mpc for a water Cherenkov detector of the size of Hyper-Kamiokande. 
Although a significant fraction of the $r$-process $\bar\nu_e$ from the BNSM dynamical ejecta emitted at earlier times are likely to be buried under the thermal neutrinos from the same BNSM, the longer emission period of the $r$-process $\bar\nu_e$ from the disk outflow can allow them to take over as the main neutrino emission source after the remnant collapses to a BH, which allows to directly probe $r$-process in BNSM, if detected.
For collapsars, we expect that the $r$-process $\bar\nu_e$ should similarly take over as the dominant component after the thermal emission subsides as in the BNSM case.

Beyond individual events, we have also computed the contribution of these sources to the diffuse neutrino background, dubbed as D$r$NB, and compared them to the expected DSNB flux. 
We found that neither source will likely produce large enough D$r$NB $\bar\nu_e$ flux at energy range currently relevant for the detection of DSNB.
We note that throughout this paper, we have not included any potential contribution from BH-NS mergers, which may also give rise to the production of $r$-process elements. 
However, as the estimation based on recent detected gravitational wave events suggests that most likely the $r$-process yields from BH-NS mergers will not dominate those from the BNSM~\cite{Chen:2021fro}, it is unlikely that including the $r$-process $\bar\nu_e$ production from BH-NS mergers will affect the conclusion derived here.

Although our results suggest that at present it is difficult to directly detect the $r$-process $\bar\nu_e$ in any near future, an (unlikely) detection can, however, provide independent and a direct probe to the $r$-process condition in BNSMs and collapsars. 
We also end with a positive note that the calculation presented in this work can be useful for examining the detailed annihilation imprints of the $r$-process $\bar\nu_e$ on the high-energy neutrino spectrum and flavor composition from collapsars~\cite{Guo:2022zyl} or BNSMs. 
We leave a more thorough investigation of this aspect to a future work. 

\label{sec:conclusion}

\acknowledgments
We thank Yong-Zhong Qian for useful discussions. 
M.-R.~W.\ acknowledges supports from the National Science and Technology Council under Grant No.~110-2112-M-001-050 and No.~111-2628-M-001-003-MY4, the Academia Sinica under Project No.~AS-CDA-109-M11, and Physics Division, National Center for Theoretical Sciences. G. G. is supported by the NSFC (No.~12205258) and the Natural Science Foundation of Shandong Province (No.~ZR2022JQ04). Y.Z.F is supported by NSFC (No.11921003 and No.12233011).

\appendix
\section{$\beta$-DECAY SPECTRA}
\label{appendix:beta_decay_spectra}

\begin{figure*}
    \centering
    \includegraphics[width=\textwidth]{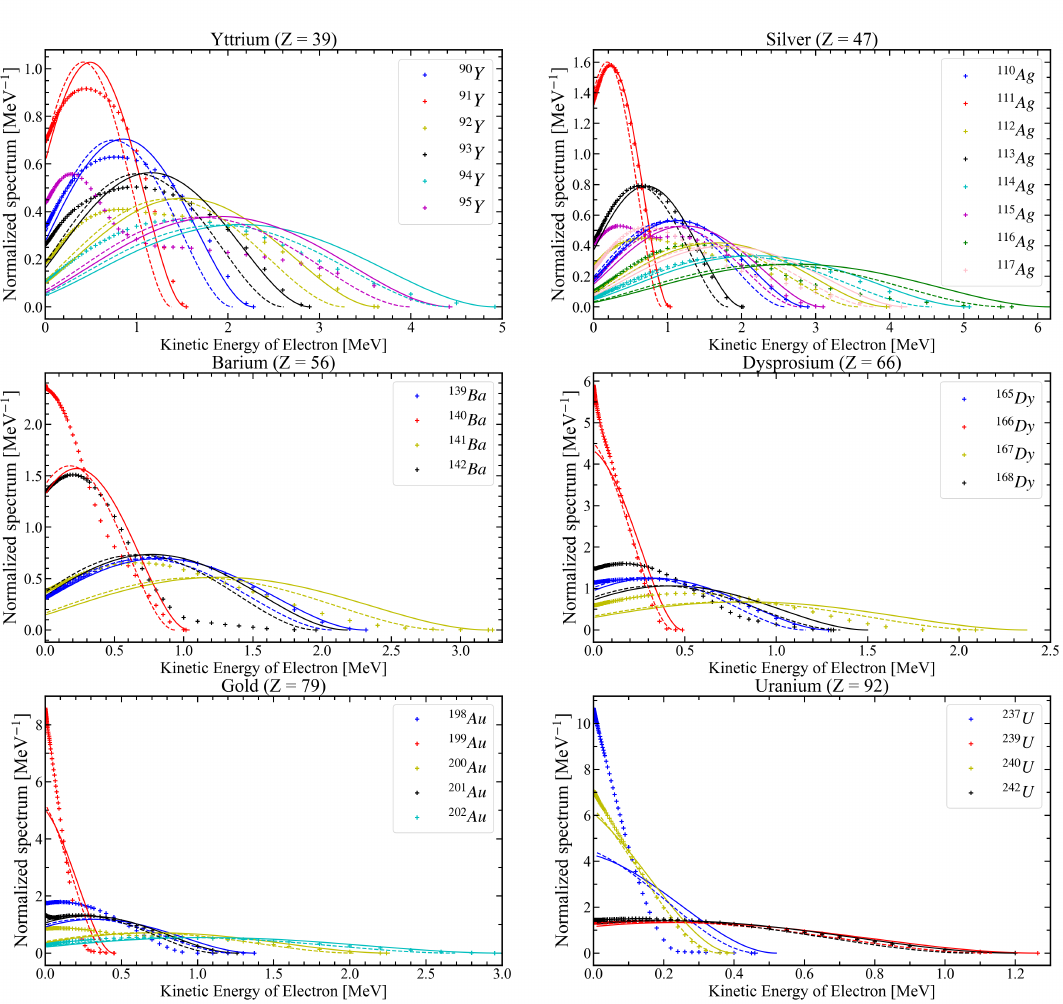}
    \caption{The comparison of the experimentally measured $\beta$-decay (normalized) spectrum for isotopes of Y, Ag, Ba, Dy, Au, U (crosses) with those computed based on the approximated formula Eq.~\eqref{eq:fe}, which assumes ground state to ground state transition only. 
    The solid (dashed) lines assume $P=1.0$ ($P=0.9$), where $P$ is an effective parameter taking into account energy loss to $\gamma$-ray emission from the decay to the excited state of the daughter nucleus.
    }
    \label{fig:beta_decay}
\end{figure*}

In this appendix we compare the $\beta$-decay $e^-$ energy spectrum derived using the same assumption made for Eq.~(\ref{eq:fnuebar}) to that measured experimentally for several nuclei. 
The normalized $\beta$-decay electron spectrum that is the counterpart of the $\bar\nu_e$ spectrum given in Eq.~(\ref{eq:fnuebar}) can be similarly written down as
\begin{equation}
\begin{split}
    f_{(Z,A)}(T_e)=
     &
     K^{\prime}
     F(Z_d, T_e)({\tilde Q} - T_e)^2(T_e + m_e)\\
     &\times \sqrt{T_e^2 + 2T_e m_e},
\end{split}
\label{eq:fe}
\end{equation}
where $K^{\prime}$ is a different normalization constant such that $\int_0^{\tilde Q}dT_ef_{(Z,A)}(T_e)=1$ and $\tilde Q=P\cdot Q$ denotes the effectively reduced $Q$ value.

We show in Fig.~\ref{fig:beta_decay} the comparison of this normalized 
$e^-$ spectrum with $P=0.9$ and $1.0$ to the measured spectrum for several isotopes of yttrium ($Z = 39$), silver ($Z = 47$), barium ($Z = 56$), dysprosium ($Z = 66$), gold ($Z = 79$) and uranium ($Z = 92$)
The experimental spectra are taken from the International Commission on Radiological Protection publication 107~\cite{Eckerman:2008nuc} and are shown as crosses.
The solid and dashed line are the normalized decay spectrum from Eq.~(\ref{eq:fe}) with $P = 1.0$ and 0.9, respectively.

The comparison shows that the formula adopted here, although simple,  roughly gives rise to $e^-$ spectra similar to those measured experimentally. 
Some specific isotopes do show larger deviation, e.g., $^{95}$Y, $^{115-117}$Ag, $^{142}$Ba, and $^{237}$U.
These differences can be attributed to the large fraction (larger than $\sim 40$\%) of the decay branching to high-level excited state of the daughter nuclei. 
On the other hand, for cases where the decay dominantly goes to the ground state of the daughter nuclei, e.g., $94.9\%$, $92\%$, $85\%$ and $78\%$ for $^{110-114}$Ag, the agreement between the experimental data and that from Eq.~\ref{eq:fe} are nearly perfect. 
The decay branching to different states of the daughter ultimately depends on a number of nuclear physics properties, e.g., the overlap of the initial and final state wave functions, as well as the detailed evaluation of the shape factor. 
For $r$-process nuclei that reply on theoretical inputs to compute the detail decay schemes, we expect a large uncertainty associated with theory models. 
Therefore, for this work we simply take $P=0.9$ to evaluate the $\bar\nu_e$ spectrum as stated in the main text.

\section{ESTIMATION OF THERMAL NEUTRINO FLUX FROM BNSM POSTMERGER REMNAN}\label{appendix:thermal_neutrino}

\begin{figure}
    \centering
    \includegraphics[width=1.0\columnwidth]{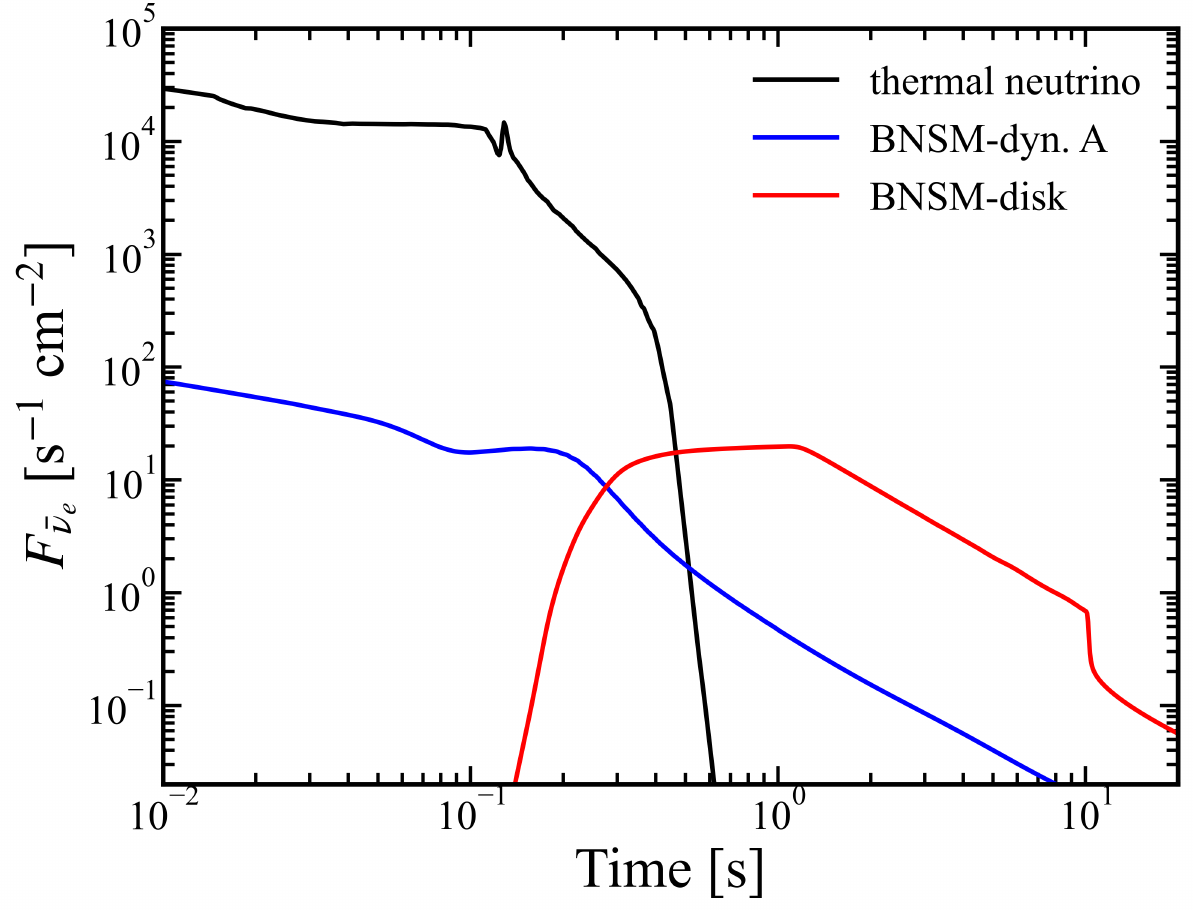}
    \caption{Comparison of the time evolution of thermal $\bar\nu_e$ flux (black solid line) computed based on the model \texttt{sym-n1-a6} from Ref~\cite{Just:2023wtj} with the $r$-process $\bar\nu_e$ emitted by dynamical ejecta~(red solid line) and disk outflow~(blue solid line) with $Y_{e, c}=0.25$ evaluated in the main text.}
    \label{fig:thermal_nue_flux}
\end{figure}

Copious amount of neutrinos are expected to be emitted from the remnant of BNSM and can affect the evolution of the system as well as the nucleosynthesis outcome (see e.g., Ref.~\cite{Fischer:2023ebq} and references therein).
The detectability of these thermal neutrinos, emitted from the HMNS before it collapsing into a BH and from the accretion disk from individual events, as well as their contribution to the diffuse neutrino background have been considered in Refs.~\cite{Kyutoku:2017wnb,Schilbach:2018bsg}. 
In this section, we take the prediction of the thermal $\bar\nu_e$ emission from 
Ref.~\cite{Just:2023wtj}, to estimate their flux and energy spectrum, and compare them with the $r$-process $\bar\nu_e$ discussed in the main text.

Ref.~\cite{Just:2023wtj} performed end-to-end simulations of BNSMs that result in delayed BH formation within $\sim 0.1-1$~s post the merger.  
Prior to the BH formation, neutrinos of all flavors are emitted from the HMNS with luminosity reaching $\sim 10^{52}-10^{53}$~erg/s for each species.  
After the HMNS collapses to BH, the emission of heavy lepton flavors gets immediately shut off, while the accretion disk continues emitting $\nu_e$ and $\bar\nu_e$ for a few hundred ms. 
Interestingly, the mass outflow rates in all models of Ref.~\cite{Just:2023wtj} follow qualitatively similar trend predicted in simulations of BH-disk systems in \cite{Fernandez:2013tya,Fahlman:2018llv}, and thus also similar to the parametrization that we adopted in the main text.
Taking the $\bar\nu_e$ energy luminosity and the average energy from the \texttt{sym-n1-a6} model of Ref.~\cite{Just:2023wtj} and assume that $\bar\nu_e$ follows the Fermi-Dirac distribution with zero chemical potential, we compute the time evolution of the $\bar\nu_e$ flux $F_{\bar\nu_e}$ at a distance of $40$~Mpc as shown in Fig.~\ref{fig:thermal_nue_flux}.  
We also compute the time-integrated neutrino energy spectrum $dN_{\bar\nu_e}/dE_{\bar\nu_e}$ emitted per source, which is shown in Fig.~\ref{fig:number_flux_spectrum} and used in the calculation of the diffuse neutrino flux in Appendix~\ref{appendix:neutrino_spectra}.

The comparison of the thermal neutrino flux to the $r$-process $\bar\nu_e$ fluxes from our BNSM dynamical ejecta~(BNSM-dyn.~A) and disk outflow~(BNSM-disk) with $Y_{e,c} = 0.25$ clearly shows that the majority of the $r$-process $\bar\nu_e$ from the dynamical ejecta as well as the early $\bar\nu_e$ emission from the disk ejecta are buried under the much large flux of thermal $\bar\nu_e$. 
However, at times $\gtrsim \mathcal{O}(1)$~s after the thermal emission shuts off, the $r$-process $\bar\nu_e$ from the disk ejecta becomes the dominating component as expected.

\section{DSNB CALCULATION AND $\bar\nu_e$ SPECTRA AT DIFFERENT SOURCES}
In this appendix we present the detail information that we used to compute the DSNB flux. 
In addition, we compare the source $\bar\nu_e$ spectra from CCSNe, BNSMs, and collapsars. 

We adopt the comoving CCSN rate $R_{\rm CCSN}$ determined by the SFR and the IMF from~\cite{Horiuchi:2008jz}
\begin{equation}
    R_{\rm CCSN} = \dot{\rho}_{\ast}(z) \frac{\int^{50}_8 dM \psi(M)}{\int^{100}_{0.1}dM M\psi(M)},
    \label{eq:sn_rate}
\end{equation}
where $M$ is zeroth-age-main-sequence mass of a star, and $\psi(M)$ is the IMF. 
Taking Salpeter's IMF~\cite{Salpeter:1955it}, the ratio of the two integrals on the right hand side equals 0.0070$/{\rm M}_{\odot}$. 
For the SFR, we assume an analytical formula $\dot{\rho}_{\ast}(z)$ described as a continuous broken power law given by~\cite{Yuksel:2008cu}
\begin{equation}
    \dot{\rho}_{\ast}(z) = 
    \dot{\rho}_{0}[(1 + z)^{a\eta} + (\frac{1 + z}{B})^{b\eta} + (\frac{1 + z}{C})^{c\eta}]^{1/{\eta}},
    \label{eq:sfr}
\end{equation}
where 
$\dot{\rho}_{0} \simeq$ 0.0178 $M_{\odot}$ yr$^{-1}$ Mpc$^{-3}$, and $a$ = 3.4 ($b = -0.3$, $c = -3.5$) is the logarithmic slope of the low (intermediate, high) red shift regimes. The red shift break constants $B$ and $C$ at $z_1=1$ and $z_2=4$ are defined as $B = (1 + z_1)^{1-a/b}$ and $C = (1 + z_1)^{(b - a)/c}(1 + z_2)^{1-b/c}$, while $\eta \simeq -10$ to smooth the transition.

For the source $\bar\nu_e$ spectrum $dN_{\bar\nu_e}/dE_{\bar\nu_e}$ for SNe, we take an approximated Fermi-Dirac form as~\cite{Beacom:2010kk}
\begin{equation}
     dN_{\bar\nu_e}/dE_{\bar\nu_e}= E_{\bar\nu_e, {\rm tot}} \times \frac{120}{7\pi^4} \frac{E_{\bar\nu_e}^2}{T^4} \frac{1}{e^{E_{\bar\nu_e}/T} + 1},    
    \label{eq: sn_spectrum}
\end{equation}
where the nominal total energy $E_{\bar\nu_e, {\rm tot}} \simeq 3 \times 10^{58}$~MeV, and the effective $\bar\nu_e$ temperature $T = 5$~MeV. 

\begin{figure}
    \centering
    \includegraphics[width=1.0\columnwidth]{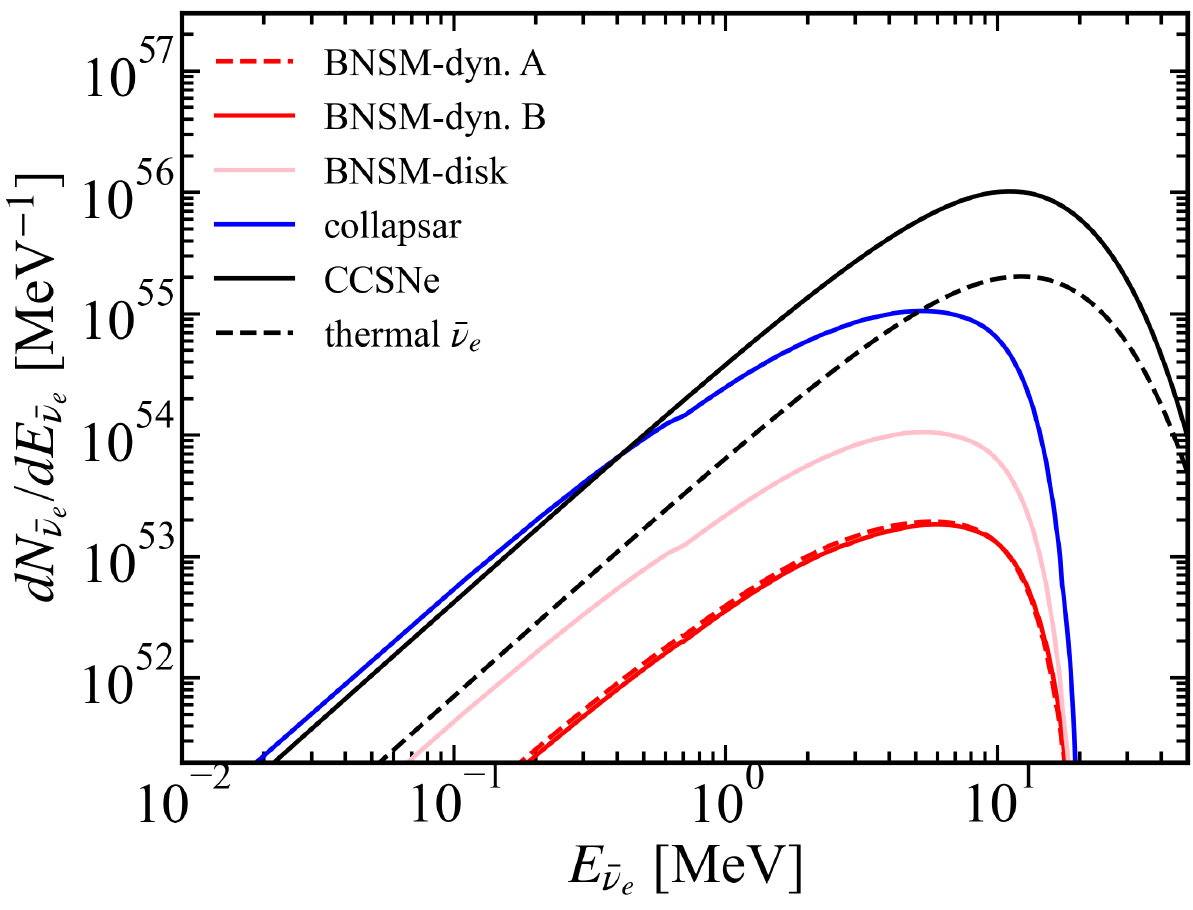}
    \caption{
    The $\bar\nu_e$ emission energy spectrum $dN_{\bar\nu_e}/dE_{\bar\nu_e}$ from a single event used for the calculation of the diffuse flux for all the considered models.
    }
    \label{fig:number_flux_spectrum}
\end{figure}

Figure~\ref{fig:number_flux_spectrum} shows the comparison of all the considered source spectra used in Sec.~\ref{sec:diffuse_neutrino}. 
Clearly, $\bar\nu_e$ from SNe are typically much larger than the $r$-process $\bar\nu_e$ from BNSM in all energy ranges.
However, for collapsars, their $\bar\nu_e$ emission at low energy $\lesssim 3$~MeV may be comparable to or higher than the SN $\bar\nu_e$, due to a potentially large amount of $r$-process enriched outflow.

\label{appendix:neutrino_spectra}

\newpage

\bibliography{ref}

\end{document}